\documentclass[preprint]{elsarticle}
\usepackage{amsmath}
\usepackage{graphicx}
\usepackage{enumerate}
\usepackage{natbib}
\usepackage{appendix}
\begin{document}

\bibpunct{(}{)}{;}{a}{,}{;} 
\title{Parameter Estimation for Multivariate Diffusion Systems}
\author{Melvin M. Varughese\footnote{Email: melvin.varughese@uct.ac.za.}}
\address{Department of Statistical Sciences, University of Cape Town, Rondebosch, Cape Town 7707, South Africa.}

\begin{abstract}
Diffusion processes are widely used for modelling real-world phenomena. Except for select cases however, analytical expressions do not exist for a diffusion process' transitional probabilities. It is proposed that the cumulant truncation procedure can be applied to predict the evolution of the cumulants of the system. These predictions may be subsequently used within the saddlepoint procedure to approximate the transitional probabilities. An approximation to the likelihood of the diffusion system is then easily derived. The method is applicable for a wide-range of diffusion systems - including multivariate, irreducible diffusion systems that existing estimation schemes struggle with. Not only is the accuracy of the saddlepoint comparable with the Hermite expansion - a popular approximation to a diffusion system's transitional density - it also appears to be less susceptible to increasing lags between successive samplings of the diffusion process. Furthermore, the saddlepoint is more stable in regions of the parameter space that are far from the maximum likelihood estimates. Hence, the saddlepoint method can be naturally incorporated within a Markov Chain Monte Carlo (MCMC) routine in order to provide reliable estimates and credibility intervals of the diffusion model's parameters. The method is applied to fit the Heston model to daily observations of the S\&P 500 and VIX indices from December 2009 to November 2010.
\end{abstract}

%
%
\maketitle

\section{Introduction}
Diffusion processes are continuous-time, continuous-space stochastic processes that have proven to be natural modelling frameworks for many real world phenomena. Over an infinitesimal interval $dt$, the evolution of a multivariate diffusion process $\phi_t^*$ is represented by the following, possibly time inhomogeneous, stochastic differential equation (SDE):
\begin{equation}
d\phi_t^*=\mu\left(\phi_t^*,t;\theta\right)dt+\sigma\left(\phi_t^*,t;\theta\right)dB_t,
\label{SDE}
\end{equation}
where $\phi_t^*=\left(\phi_i\right)_{i=1,...,m}$; $\theta=\left(\theta_i\right)_{i=1,...,p}$ is the parameter vector; $\mu\left(\phi_t^*,t;\theta\right)=\left(\mu_i\right)_{i=1,...,m}$; $\sigma^2\left(\phi_t^*,t;\theta\right)=$  $\left(\sigma_{ij}\right)_{i,j=1,...,m}$ with $\sigma^2\left(\phi_t^*,t;\theta\right)=\sigma\left(\phi_t^*,t;\theta\right)^T\sigma\left(\phi_t^*,t;\theta\right)$ and $B_t$ is an $m$-dimensional vector of independent Brownian motions. The $m$-dimensional vector $\phi_t^*$ represents a set of state variables which characterizes the diffusion system through time. The assumption that the $m$ Brownian motions are independent does not lead to any loss of generality since allowance is made for off-diagonal terms within the diffusion matrix $\sigma^2\left(\phi_t^*,t;\theta\right)$. Within this framework, the primary focus is on estimating the parameter vector $\theta$ from discretely-sampled data.

The drift vector $\mu\left(\phi_t^*,t;\theta\right)$ and the diffusion matrix $\sigma^2\left(\phi_t^*,t;\theta\right)$ characterize the evolution of $\phi_t^*$. Their individual elements are defined as:
\begin{equation}
\mu_i = \lim_{\Delta t\rightarrow 0}\frac{\mbox{E}[\Delta \phi_i|\phi_t^*]}{\Delta t}, \;\;
\sigma_{ij} = \lim_{\Delta t\rightarrow 0}\frac{\mbox{Cov}[\Delta\phi_i,\Delta\phi_j|\phi_t^*]}{\Delta t}. \nonumber
\end{equation}
Over an infinitesimally small interval, the diffusion system is distributed as follows:
\begin{equation}
\phi_{t+\Delta t}^*-\phi_t^*\sim \mbox{Normal}\left(\mu\left(\phi_t^*,t;\theta\right)\Delta t, \sigma^2\left(\phi_t^*,t;\theta\right)\Delta t \right).
\label{incdist}
\end{equation}
That is, the diffusion system has a multivariate-normal distribution characterized by the drift vector and diffusion matrix over any infinitesimal interval. Since diffusion processes are Markovian, equation (\ref{incdist}) may be used to derive the likelihood for {\em{continuously}} sampled diffusion paths. However with {\em{discretely}} sampled diffusion paths, statistical inference is considerably more challenging. This is because the distribution of the diffusion increments over discretely sampled time points is often unknown.

Instead of representing a multivariate diffusion system $\phi_t^*$ as a stochastic differential equation, one may instead focus on the Kolmogorov forward equation, which dictates the evolution of its probability density function $p(\phi_t^*)$. This is given by:
\begin{eqnarray}
\frac{\partial p(\phi_t^*)}{\partial t}&=&-\sum_{i=1}^m\frac{\partial}{\partial \phi_i}\left[\mu\left(\phi_t^*,t;\theta\right)p(\phi_t^*)\right]\nonumber\\
&+&\frac{1}{2}\sum_{i=1}^m\sum_{j=1}^m \frac{\partial ^2}{\partial \phi_i \partial \phi_j} \left[\sigma_{ij}(\phi_t^*,t;\theta)p(\phi_t^*)\right].
\label{Fokker}
\end{eqnarray}
This is also known as the Fokker-Planck equation. Except where required, the dependence of the drift vector and diffusion tensor on the parameter vector shall be suppressed within the notation. Since a diffusion process is Markovian, the likelihood of a diffusion system sampled at discrete time points $(t_1, t_2,..., t_N)$ is given by:
\begin{equation}
L(\theta)=p(\phi_{t_1}^*)\prod_{i=2}^N p\left(\phi_{t_i}^*|\phi_{t_{i-1}}^*\right).
\label{like}
\end{equation}
Asymptotically, for large $N$, the term $p(\phi_{t_1}^*)$ may be ignored whilst the transitional probability distribution $p(\phi_{t_i}^*|\phi_{t_{i-1}}^*)$  is the solution to equation (\ref{Fokker}) at time $t_i$ with the boundary condition that $p(\phi_t^*)$ is given by the Dirac delta function centered around $\phi_{t_{i-1}}^*$  at time $t_{i-1}$. The likelihood function is central to many inference procedures: it enables us to derive parameter estimates, confidence intervals and to conduct hypothesis tests. Unfortunately, except for a few special cases, equation (\ref{Fokker}) (and hence also the likelihood) is analytically intractable.

The inability to solve equation (\ref{Fokker}) is an impediment to statistical inference. This may be circumvented by attempting to match, by choice of parameters, characteristics of the sampled path with characteristics of the diffusion model. For example, one may choose to estimate the instantaneous means and variances using the corresponding sample moments of the differenced data \citep{Gallant1997,Ragwitz01}. Alternatively one may employ Bayesian imputation to augment the observed data so that the diffusion increments are approximately normally distributed \citep{Roberts01}.

Since likelihood based methods tend to give more precise parameter estimates than method of moments estimators \citep{Hurn07}, we instead seek an approximation to the transitional probability distribution of the diffusion process. Monte Carlo methods may be used to approximate the likelihood \citep{Kleinhans07,Durham02}. Alternatively, equation (\ref{Fokker}) could be solved numerically.  \citet{Woj00} discretized the spatial domain and solved the partial differential equation numerically at each point on the lattice. The finite-difference method discretizes the time domain, taking advantage of the fact that over an infinitesimally small time period, the diffusion process is normally distributed \citep{Wehner87}. \citet{Huang12} developed a quasi-maximum likelihood estimator that approximates the first two conditional moments using a Wagner-Platen approximation. The resulting normal distribution can be used to approximate the transitional probability.

Another possibility is to approximate the transitional probability distribution by a closed form analytic function; for example, a Hermite polynomial expansion \citep{Ait2002}. This method was shown by Ait-Sahalia to be superior to many of the competing methods - both in terms of the accuracy of the transitional distribution approximation as well as the speed of the algorithm. \citet{Stramer10} created an MCMC procedure based on the Hermite approximation which could allow for measurement errors in the diffusion process.

It must be stressed that the Hermite approximation is only applicable for reducible diffusion processes. A diffusion process
$X$ is reducible if there exists a one-to-one transformation $Y=h(X,\theta)$ such that the covariance function of $Y$ is the identity matrix. Though all univariate diffusion processes are reducible, only some multivariate diffusions share this property. \citet{Ait2008} extended the method to irreducible, multivariate diffusions, but not only is the procedure more difficult to implement, there is also a reduction in the accuracy of the closed-form approximation to the transitional density. Furthermore, the Hermite approximation does not in general integrate to one. Indeed, for parameter values far from the maximum likelihood estimates, the normalizer can be very far from one. This often prevents convergence when applying the Hermite approximation within an MCMC setting \citep{Stramer10}. Consequently, without modification, the resulting MCMC credibility intervals often suffer from considerable undercoverage.

It is proposed that the transitional probability may rather be estimated by a saddlepoint approximation \citep{Daniels54}. The saddlepoint approximation is an algebraic expression based on a random variable's cumulant generation function (CGF). In cases where the first few moments of a random variable is known but the corresponding probability density is difficult to obtain, the saddlepoint approximation to the density can be calculated. The tails of a saddlepoint approximation are more accurate than those of a Edgeworth-expansion \citep{Barndorff99}. Saddlepoint methods have already been used to approximate the transition densities of diffusions \citep{Ait2006,Preston11}. Preston and Wood apply the saddlepoint approximation to the CGF of a truncated small-time sample-path expansion whilst \citet{Ait2006} apply the saddlepoint to a truncated expansion, in small-time, of the characteristic function of the transition density. 

The saddlepoint approximation will be applied to a truncated expansion of the CGF of $\phi_t^*$ with respect to the CGF parameters. In contrast to the aforementioned small-time methods, the focus is instead on predicting the evolution of the diffusion system's cumulants through time, which may be achieved with the cumulant truncation procedure \citep{Whittle57,Gillespie07}. The cumulant truncation procedure may also be applied to multivariate diffusion processes \citep{Me08,Me11}. These predicted cumulants can be subsequently substituted within a saddlepoint to approximate the transitional densities. After fitting a diffusion model, one may subsequently test for model misspecification of the SDE \citep{Zhang2012}.

In Section 2, the saddlepoint approximation to the transitional probability distribution of the diffusion process is introduced. In Section 3, an MCMC algorithm which uses this approximation for parameter estimation is introduced. Since \citet{Ait2002} demonstrated that the Hermite approximation is superior to a number of alternative approximations, we compare the saddlepoint and the Hermite approximation in Section 4. The two methods are tested on univariate diffusion processes. The Heston model (an irreducible, multivariate diffusion process) is fitted in Section 5 to the S\&P 500 and VIX indices. This is followed by a discussion of the various results and some conclusions being drawn.

\section{Approximating the transitional probability distribution}
\label{sec:Sadd}
\subsection{The saddlepoint approximation}
In cases where the distribution of a random vector $\mathbf{X}^T=(X_1,X_2,...,X_m)$ is unknown, the saddlepoint method provides an algebraic approximation based on $\mathbf{X}$'s cumulant generating function. Let \begin{small}$\Lambda^T=(\upsilon_1,..., \upsilon_m)$ \end{small} be the parameters for $\mathbf{X}$'s moment generating function (MGF) $M(\Lambda)=E\left[\exp[\upsilon_1X_1+...+\upsilon_mX_m]\right]$. Also, let $K(\Lambda)=\ln M(\Lambda)$ denote the cumulant generating function (CGF). Assuming the MGF exists in an open neighbourhood around the origin \citep{Daniels54}, the leading order $m$-dimensional saddlepoint approximation is given by:
\begin{small}
\begin{equation}
	f(\mathbf{x})=(2\pi)^{-m/2}\left|\nabla^2K(\Lambda)\right|^{-1/2}\exp\left\{K(\Lambda)-\Lambda^T\mathbf{x}\right\},
\label{Msaddle}
\end{equation}
\end{small}
\noindent where $\mathbf{x}$ is an $m$-dimensional vector and $\nabla^2K(\Lambda)$ represents the $m\times m$ Hessian matrix for $K(\Lambda)$ and
\[
	\nabla K(\Lambda)=\mathbf{x}.
\]
Except for some select cases such as the Normal and the Gamma distribution, substituting the true CGF within equation (\ref{Msaddle}) does not simplify to the true distribution, but rather to an approximation thereof. 

In many cases, the CGF $K(\Lambda)$ is unknown, but it may be approximated. Given the first $n$ cumulants $(\kappa_1, \kappa_2, ..., \kappa_n)$ for a univariate random variable $X$, $K(\theta)=E[\exp\{\theta X\}]$ may be approximated by:
\begin{equation}
	K(\theta)\approx\sum_{i=1}^n{\frac{\theta^n}{n!}\kappa_n}.
	\label{acgf}
\end{equation}
In the univariate case, the saddlepoint is consequently given by:
\begin{small}
\begin{equation}
\begin{split}
&f_n(x)=\left(2\pi\sum_{i=0}^{n-2}\frac{\kappa_{i+2}\theta_0^i}{i!}\right)^{-\frac{1}{2}}\exp\left[\sum_{i=1}^n \frac{\kappa_i \theta_0^i}{i!}-\theta_0 x\right],\\
&\mbox{ where: } x=\sum_{i=0}^{n-1} \frac{\kappa_{i+1}\theta_0^i}{i!}.
\end{split}
\label{saddle}
\end{equation}
\end{small}
The approximation within (\ref{saddle}) is two-fold as not only is the saddlepoint an approximation to the true density, but the saddlepoint is now based on a truncated approximation to the true CGF. For a diffusion process, the approximate evolution of the cumulants (and hence also the approximation evolution of equation (\ref{acgf})) can be calculated through the cumulant truncation procedure.

\subsection{Applying the cumulant truncation procedure to a diffusion process}
Consider a diffusion process $\phi_t^*=\left(\phi_i\right)_{i=1,...,m}$ with MGF \linebreak $M(\Lambda,t)=E\left[\exp[\upsilon_1\phi_1(t)+\upsilon_2\phi_2(t)+...+\upsilon_m\phi_m(t)]\right]$ and CGF \linebreak $K(\Lambda,t)=\ln M(\Lambda,t)$. In order to calculate the likelihood given in (\ref{like}), we approximate the transitional density  $p(\phi_{t_i}^*|\phi_{t_{i-1}}^*)$ by a saddlepoint $f(\mathbf{x})$. This necessitates the calculation of an approximation to $K(\Lambda,t)$ at time $t_i$ given that the diffusion process is equal to $\phi_{t_{i-1}}^*$ at time $t_{i-1}$. 

Let $\Lambda^*=(\partial / \partial\upsilon_1,..., \partial / \partial\upsilon_m)$. Like the partial differential equation for the probability density (given in (\ref{Fokker})), there is a corresponding partial differential equation for the MGF $M(\Lambda,t)$, the solution of which enables us to predict the evolution of the system's cumulants. The MGF obeys the following relation (see Appendix):
\begin{equation}
\frac{\partial M(\Lambda,t)}{\partial t}=\bigg[\sum_{i=1}^m \upsilon_i\mu_i(\Lambda^*,t)
+\frac{1}{2}\sum_{i=1}^m\sum_{j=1}^m \upsilon_i\upsilon_j\sigma_{ij}(\Lambda^*,t)\bigg]M(\Lambda,t).
\label{MGF}
\end{equation}
Though the terms $\mu_i(\Lambda,t)$ and $\sigma_{ij}(\Lambda,t)$ denote functions of $\Lambda$ and $t$, the terms $\mu_i(\Lambda^*,t)$ and $\sigma_{ij}(\Lambda^*,t)$ represent differential operators that act on the MGF $M=M(\Lambda,t)$. So, for example, if we have a 2-dimensional diffusion system with drift term $\mu_1(\phi_1,\phi_2)=a\phi_1-b\phi_1^2+c\phi_1\phi_2$, then $\mu_1\left(\frac{\partial}{\partial \upsilon_1},\frac{\partial}{\partial \upsilon_2}\right)=a\frac{\partial}{\partial \upsilon_1}-b\frac{\partial^2}{\partial \upsilon_1^2}+c\frac{\partial^2}{\partial \upsilon_1\partial \upsilon_2}$ is a differential operator and $\mu_1\left(\frac{\partial}{\partial \upsilon_1},\frac{\partial}{\partial \upsilon_2}\right)M=a\frac{\partial M}{\partial \upsilon_1}-b\frac{\partial^2 M}{\partial \upsilon_1^2}+c\frac{\partial^2 M}{\partial \upsilon_1\partial \upsilon_2}$ represents the action of the operator on the MGF $M(\Lambda,t)$. The MGF may be used to characterize the evolution of the diffusion system through time. Unfortunately, as with equation (\ref{Fokker}), equation (\ref{MGF}) is generally analytically intractable.

The cumulant truncation procedure enables us to convert the partial differential equation for the MGF into a system of ordinary differential equations that also depends on the system parameters $\theta$. The resulting set of equations proves to be far easier to solve numerically than the original partial differential equation. Though, in principle, one can derive a system of ordinary differential equations directly from equation (\ref{MGF}), the cumulant truncation procedure is instead based on the corresponding partial differential equation for the CGF $K(\Lambda,t) = \ln M(\Lambda,t)$. This is because $K(\Lambda,t)$ can be expressed as a series expansion of the diffusion system's cumulants.
\begin{equation}
K=\sum_{r_1\geq0}\sum_{r_2\geq0}...\sum_{r_m\geq0}\frac{\left(\prod_{i=1}^m\upsilon_i^{r_i}\right)\kappa_{r_1,r_2,...,r_m}(t)}{\prod_{i=1}^m r_i!},
\label{CGF}
\end{equation}
where:
\begin{small}
\[
\kappa_{r_1,...,r_m}(t)=\left\{
																	 \begin{array}{ll}
																	 		\mbox{E}[\phi_i] & (r_i=1, r_j=0; j\neq i) \\
																	 		\mbox{E}\left[\prod_{i=1}^m\left(\phi_i-E[\phi_i]\right)^{r_i}\right] & (4>\sum_{j=1}^m r_j>1).
																	 \end{array}\right.
\]
\end{small}
In order to be able to predict the evolution of the diffusion system's cumulants, we need to derive a partial differential equation governing the evolution of $K(\Lambda,t)$.  Since $K(\Lambda,t) = \ln M(\Lambda,t)$, we have:
\begin{equation}
\frac{1}{M}\frac{\partial M}{\partial x}=\frac{\partial K}{\partial x},
\label{rel}
\end{equation}
where $M$ and $K$ depend on some variable $x$. By repeatedly applying the above relation, it is possible to derive the corresponding partial differential equation for the CGF from the partial differential equation for the MGF  given by equation (\ref{MGF}). The general form of this equation cannot be shown, but rather must be independently derived for each system of interest. The CGF partial differential equation is derived for selected examples within Sections \ref{sec:Comp} and \ref{sec:Ex}.

By substituting the series expansion in (\ref{CGF}), truncated to order $n$, within the partial differential equation for $K(\Lambda,t)$ and subsequently matching the cumulant coefficients \begin{small}$\left\{\left(\prod_{i=1}^m \upsilon_i^{r_i} \bigg/ \prod_{i=1}^m r_i! \right):r_i=0,1,2,...; 1\leq i \leq m \right\}$\end{small}, it is possible to derive a system of ordinary differential equations for the centered moments or cumulants. Consequently, by solving this system of differential equations, it is possible to predict the approximate evolution of the cumulants for each proposed parameter vector $\theta$. Throughout the paper this analysis was performed using MATHEMATICA v6.0. These predictions are accurate across a wide range of parameter values \citep{Me09}. This is key, as we can subsequently use the predicted cumulants within the saddlepoint to approximate $p(\phi_{t_i}^*|\phi_{t_{i-1}}^*)$ and hence derive the approximate likelihood at $\theta$. The parameter estimates are taken to be the values that maximize this approximate likelihood, which may be estimated by a Markov Chain Monte Carlo (MCMC) procedure.


\section{A modified MCMC parameter estimation algorithm}
\label{sec:MCMC}
In this section, a parameter estimation algorithm for a diffusion system is proposed. The algorithm uses the saddlepoint approximation introduced in Section \ref{sec:Sadd} to recreate the transitional probability distribution. This is subsequently used to approximate the likelihood function of the diffusion system. The approximate likelihoods enable us to explore the parameter space of the diffusion system with a modified MCMC procedure. The algorithm is described below.

\begin{enumerate}
\item Apply the cumulant truncation procedure to the diffusion system thus deriving a system of ordinary differential equations that describe the cumulants' evolution. (Note these differential equations will depend on the model parameters $\theta$.)
\item Choose an arbitrary, starting set of parameter values $\theta_0$. Set $\theta_{old} = \theta_0$.
\item Propose a jump from the old set of parameter values $\theta_{old}$ to a new set of parameter values, $\theta_{new} = \theta_{old} + \Delta\theta$ using a suitably chosen proposal distribution $q(\theta_{new}|\theta_{old})$.
\item Calculate the likelihoods $L(\theta_{old})$ and $L(\theta_{new})$. $L(\theta)$ is calculated as follows:
	\begin{enumerate}[i.]
	\item Set the likelihood, $L(\theta) = 1$.
	\item Set $i$ to 1.
	\item Use the system of ordinary differential equations from Step 1 together with the data values observed at time $t_i$, $\phi_{t_i}^*=(\phi_1,\phi_2,...,\phi_m)$ to predict the values of the cumulants at time $t_{i+1}$.
	\item Given the data $\phi_{t_i}^*$  at time $t_i$, the transitional probability distribution at time $t_{i+1}$ can be approximated by a saddlepoint approximation \break $f_n(\phi_{t_{i+1}}^*|\phi_{t_i}^*)$ after substituting for the cumulants derived from step iii.
	\item Set the likelihood $L(\theta)=L(\theta)\times f_n(\phi_{t_{i+1}}^*|\phi_{t_i}^*)$. This is an implementation of equation (\ref{like}). 
	\item Set $i=i+1$ and if $i<N$ (where $N$ is the number of data points) go to step iii.
	\end{enumerate}
\item	Calculate the acceptance ratio $R$. This is given by:
\[R=\frac{L(\theta_{new})}{L(\theta_{old})}\frac{\pi(\theta_{new})}{\pi(\theta_{old})}\frac{q(\theta_{old}|\theta_{new})}{q(\theta_{new}|\theta_{old})},\]
where $\pi(\theta)$ is the prior density. We then accept the proposed parameter values  $\theta_{new}$ with probability $\min(1,R)$.
That is, set $\theta_{new} = \theta_{old}$ with probability $1-\min(1,R)$. Otherwise leave $\theta_{new}$ unchanged.
\item Go back to step 3.
\end{enumerate}
After a suitably chosen burn-in period, the MCMC chains sample from the posterior distribution of the diffusion parameters. Hence, in addition to using the medians of the chains as estimates of the diffusion parameters, the $\alpha/2$-th and $(1-\alpha/2)$-th percentile may be used to construct $(100-\alpha)$\% credibility intervals.

\section{A comparative study of the Saddlepoint and Hermite methods}
\label{sec:Comp}
Since the Hermite approximation \citep{Ait2002} is a popular method for estimating the parameters of a diffusion system, we compare its performance against the saddlepoint. In order to judge the relative accuracies of the saddlepoint and Hermite procedures, the true transitional probability distribution is required as a baseline. Unfortunately, the diffusion models for which the transitional distribution is known are few. Amongst, diffusion processes with analytical, non-normal transitional distributions, the Cox-Ingersoll-Ross (CIR) process and Geometric Brownian motion are arguably the most well known. Neither the saddlepoint nor the Hermite approximation in general integrate to one and the approximations are not normalized when performing the comparisons in this section.

Within this section, the CIR process is analyzed didactically: the analysis is meant to illustrate the general derivation of the saddlepoint approximation to a transitional distribution. This is followed by a study of the relative accuracy, for the CIR process, of the Hermite and saddlepoint approximations to the transitional distribution. The section concludes by comparing MCMC implementations for the Hermite and saddlepoint procedures for both the CIR process as well as Geometric Brownian motion.

\subsection{Example: Deriving the saddlepoint approximation for the CIR process}
The CIR process is commonly used to model financial data such as short-term interest rates. The model may be represented as follows:
\begin{equation}
dX_t=b(\mu-X_t)dt+\sigma\sqrt{X_t}dB_t.
\label{CIR}
\end{equation}
The CIR process is mean-reverting. Furthermore, provided $2b\mu>\sigma^2$, the CIR process is always positive. This is advantageous as many real-world phenomena are positive and display mean reversion. The probability distribution of the process obeys the following partial differential equation:
\begin{equation}
\frac{\partial p(X_t)}{\partial t}=-\frac{\partial}{\partial X_t}\left[b(\mu-X_t)p(X_t)\right]+\frac{1}{2}\frac{\partial^2}{\partial X_t^2}\left[\sigma^2X_tp(X_t)\right].
\label{CIRpdf}
\end{equation}
An analytical solution to equation (\ref{CIRpdf}) exists \citep{Cox85}. Hence it is possible to derive an analytical expression for the transitional distribution which may be subsequently compared with the Hermite and the saddlepoint approximations. This enables us to compare the relative accuracies of the two approximation schemes. 

The first step of the cumulant truncation procedure is to derive the evolution of the moment generating function through time. This requires the drift and diffusion terms of the process. These are given by:
\[
\mu(X_t,t)=b(\mu-X_t);\;\sigma(X_t,t)=\sigma\sqrt{X_t}.
\]
By substituting the drift and diffusion terms within equation (\ref{MGF}), we derive the partial differential equation for the MGF $M(\lambda,t)=E\left[\exp[\lambda X_t]\right]$:
\begin{equation}
\frac{\partial M}{\partial t}=\lambda b\mu M-\left[\lambda b - \frac{1}{2}\lambda^2\sigma^2\right]\frac{\partial M}{\partial \lambda}.
\label{CIRMGF}
\end{equation}
By dividing both sides of equation (\ref{CIRMGF}) by $M(\lambda,t)$ and applying equation (\ref{rel}), it is possible to derive the differential equation for the CGF:
\begin{equation}
\frac{\partial K}{\partial t}=\lambda b \mu - \left[\lambda b -\frac{1}{2}\lambda^2\sigma^2\right]\frac{\partial K}{\partial \lambda}.
\label{CIRCGF}
\end{equation}
In most cases the partial differential equation for the CGF will be analytically intractable. Under such scenarios, one may instead substitute a truncated expansion of the CGF in place of the full CGF. As the order of the expansion increases, the accuracy of the approximation tends to increase \citep{Me09}. Suppose we approximate $K(\lambda,t)$ by a fourth-order expansion of the CGF:
\begin{equation}
K(\lambda,t)\approx 1+\lambda \kappa_1(t)+\frac{\lambda^2}{2}\kappa_2(t)+\frac{\lambda^3}{3!}\kappa_3(t)+\frac{\lambda^4}{4!}\kappa_4(t).
\label{CGFexp}
\end{equation}
By substituting the above expansion within equation (\ref{CIRCGF}) and subsequently equating the various coefficients of $\left\{\lambda^i : i=1,..,4\right\}$, it is possible to derive a system of ordinary differential equations that describe the evolution of the cumulants:
\begin{align}
\dot{\kappa}_1(t)&=b\left[\mu-\kappa_1(t)\right] \nonumber\\
\dot{\kappa}_2(t)&=\sigma^2\kappa_1(t)-2b\kappa_2(t) \nonumber\\
\dot{\kappa}_3(t)&=-3b\kappa_3(t)+3\sigma^2\kappa_2(t) \nonumber\\
\dot{\kappa}_4(t)&=-4b\kappa_4(t)+6\sigma^2\kappa_3(t). \nonumber
\end{align}
Note that there is a trade-off as the order of the approximation to the CGF increases. Not only does the approximation's accuracy increase, but the complexity of the system of equations also increases, which will lead to a rise in computing time. A fourth order approximation is chosen as it seems to be a good balance between accuracy and speed.

Boundary conditions must be specified in order to make the solution of the above system of equations unique. When predicting the cumulants at time $t_i$, the boundary conditions are taken to be: $\kappa_1(t_{i-1})=\phi_{t_{i-1}}$ , $\kappa_2(t_{i-1})=0$, $\kappa_3(t_{i-1})=0$,  $\kappa_4(t_{i-1})=0$. That is, in predicting the cumulants at any future time $t_i$, we condition on the observed value of the process at time $t_{i-1}$. This enables the system of ordinary differential equations to be solved numerically. The resulting predictions may be inserted into equation (\ref{saddle}) to give us the saddlepoint approximation to the transitional probability distribution $p(X_{t_i}|X_{t_{i-1}})$.

Since the CGF is truncated after the fourth cumulant, we shall approximate the transitional probability distribution by $f_4(x)$. From equation (\ref{saddle}) it follows that:
\begin{equation}
f_4(x)=\left[2\pi\left(\kappa_2+\kappa_3\theta_0+\frac{\kappa_4}{2}\theta_0^2\right)\right]^{-\frac{1}{2}}\exp\left[-\frac{\theta_0^2}{2}\kappa_2-\frac{\theta_0^3}{3}\kappa_3-\frac{\theta_0^4}{8}\kappa_4\right],
\label{CIRsaddle}
\end{equation}
where:
\[
x=\kappa_1+\theta_0\kappa_2+\frac{\theta_0^2}{2}\kappa_3+\frac{\theta_0^3}{6}\kappa_4.
\]
Note that $\theta_0$ is a function of $x$. A cursory glance at the above equation might suggest that $f_4(x)$ does not depend on the first cumulant. This however is incorrect since $\theta_0$ depends on $\kappa_1$.

\subsection{The relative accuracies of the transitional distribution approximations}
\begin{figure}[h!]
	\caption{The relative accuracies of the saddlepoint and Hermite approximations for the CIR process at $b=1.5$, $\mu=58$ and $\sigma^2=15$. The right panel is a plot of three curves: the true transitional probability together with the saddlepoint and Hermite approximations. The saddlepoint and Hermite approximations follow the true transitional probability distribution (a non-central Chi-squared distribution) very closely. The left panel highlights the absolute difference between the true transitional density and the two approximations. Overall, the saddlepoint approximation is more accurate}
	\begin{center}
		\includegraphics[width=2.40in]{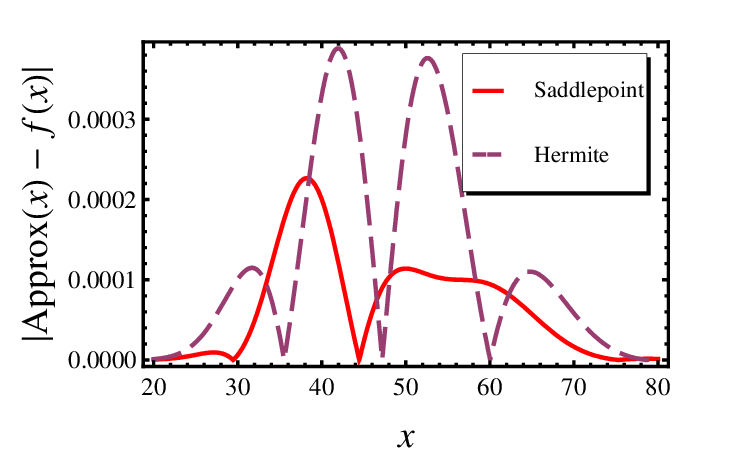}
		\includegraphics[width=2.25in]{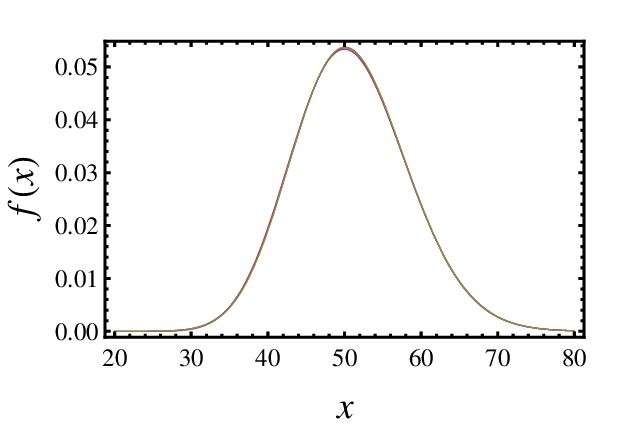}
	\end{center}
	\label{approx}
\end{figure}

The transitional distribution of the CIR process may be approximated using both the saddlepoint and Hermite approximations. Suppose the current value of the process $X_t$ is assumed to be 50 and we are interested in the probability distribution one month from now. That is, $t_i - t_{i-1} = 1/12$. Figure \ref{approx} compares the relative accuracies of the two methods, implemented in MATHEMATICA v6.0, at the parameter values:     $b = 1.5$;    $\mu = 58$;    $\sigma^2 = 15$.  On the whole, the saddlepoint approximation appears to be more accurate.

Let the Integrated Error of an approximation $\widehat{f(x)}$ be defined as:
\[
	\int_0^{\infty}{|f(x)-\widehat{f(x)}|dx}.
\]
To get a better idea of how the relative accuracy of the two approximations behaves throughout the parameter space, we plot the Integrated Error of the two models for varying values of the parameters $b$ and $\sigma^2$.
\begin{figure}[h!]
	\caption{The relative accuracies of the saddlepoint and Hermite approximations to the CIR process for varying values of $b$ (left panel) and $\sigma^2$ (right panel). Unless a parameter is varying, the values are fixed at $b=1.5$, $\mu=58$ and $\sigma^2=15$.}
	\begin{center}
		\includegraphics[width=2.36in]{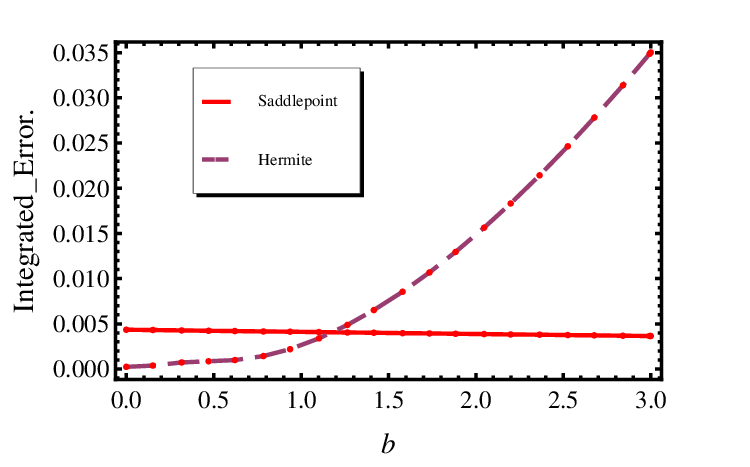}
		\includegraphics[width=2.36in]{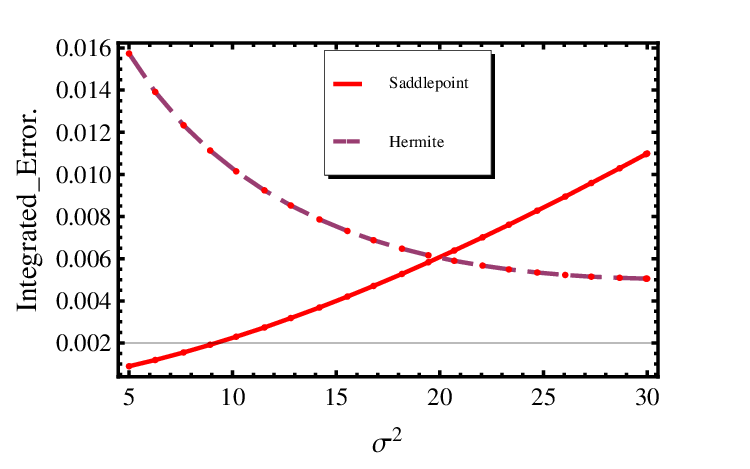}
	\end{center}
	\label{compare}
\end{figure}
One can see from Figure \ref{compare} that there are regions in the parameter space where the Hermite approximation is more accurate and regions where the saddlepoint approximation is more accurate. 

\begin{figure}[h!]
	\caption{The relative accuracies of the saddlepoint and Hermite approximations for varying $t_i-t_{i-1}$ for the CIR process at $b=1.5$, $\mu=58$ and $\sigma^2=15$. The Hermite approximation is far less accurate for larger time scales.}
	\begin{center}
		\includegraphics[width=2.74in]{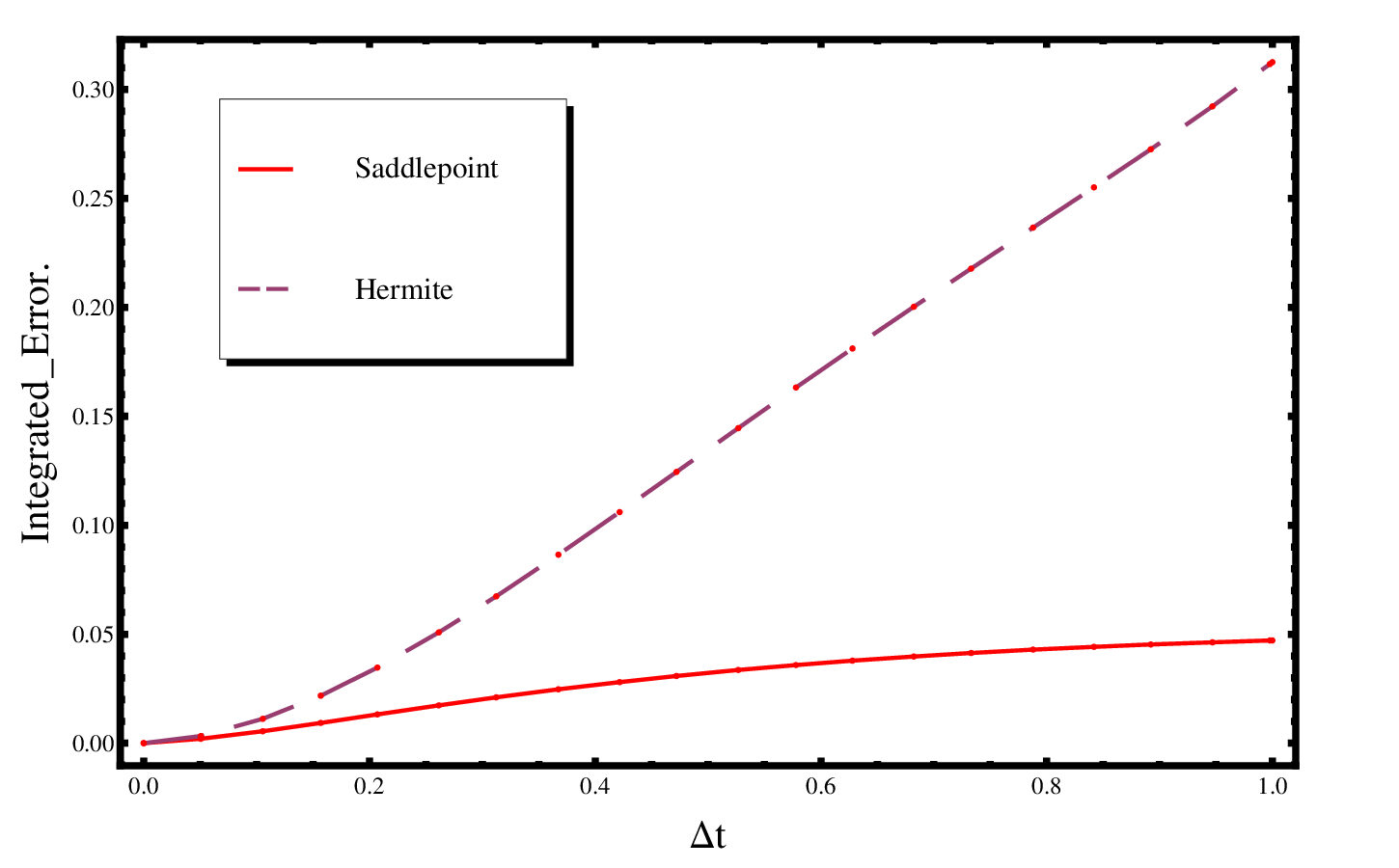}
	\end{center}
	\label{Delta}
\end{figure}

As a final point, note that unlike the saddlepoint approximation, the Hermite approximation is a small-time expansion. Consequently, the larger $t_i - t_{i-1}$ becomes, the less accurate the Hermite approximation will be. Figure \ref{Delta} shows how the Integrated Error for both the saddlepoint and the Hermite approximations changes with $t_i-t_{i-1}$. The Hermite approximation  loses accuracy more quickly for larger time scales.

This investigation has been restricted to a subset of the parameter space for the CIR model. In some sense however, the accuracy of the transitional distribution is only important in as far as it enables us to obtain precise parameter estimates for the diffusion model. With multiple parameters, the computational efficiency of the MCMC procedure makes it suitable for parameter estimation. This motivates a comparison of the modified saddlepoint MCMC algorithm described in Section \ref{sec:MCMC} with modified MCMC algorithms that uses the Hermite approximation.

\subsection{A comparison of MCMC implementations of the two approximations}
\label{sec:MCMCexamples}
Both the saddlepoint MCMC and the Hermite MCMC are judged by comparing the coverage of their resulting credibility intervals with the corresponding coverage obtained from the true MCMC. The traditional Hermite MCMC is known to fail since the Hermite approximation can explode to infinity for parameter values that are far from the maximum likelihood estimate \citep{Stramer10}. We thus run two versions of the Hermite MCMC: the traditional version that is known to fail and a modified algorithm developed by \citet{Stramer10} that avoids computation of the posterior's normalizing constant. To make the analysis more robust, the MCMC implementations are compared for both the CIR process as well as for Geometric Brownian motion. 

The empirical coverage of the credibility intervals is calculated from 100 simulated time series; each of length 40. For each of the time series, the four variants of the MCMC procedure (the saddlepoint MCMC, the two Hermite MCMCs and the true MCMC) are run and a credibility interval is calculated. We use improper, uniform priors and a normal proposal density for the parameter values when running the MCMC chains. If a parameter must be positive, any negative value sampled from the proposal distribution is thrown out and another sample is taken. The coverage of each variant of the MCMC procedure is estimated as the proportion of the 100 credibility intervals that contain the true parameter values i.e. the parameter values used to simulate the time series. In calculating the credibility intervals, an MCMC chain of length 20,000 was run and the first 10,000 steps were trimmed as part of the burn-in period. 90\% credibility intervals will correspond to the 5-th and 95-th percentiles of the resulting chains. It is desirable that the credibility intervals of a proposed MCMC implementation match as closely as possible the observed coverages of the true MCMC. 

Despite the computational efficiency of both the saddlepoint and the two versions of the Hermite procedures, a study of the coverage of their corresponding implementations is computationally intensive. 100 MCMC chains, each of length 20,000, must be run for the three MCMC implementations. This restricts us to comparing the coverage of their credibility intervals for a single set of parameter values. 

In order to make our analysis as robust as possible, we chose to compare the coverage of the credibility intervals at points in the parameter space where the saddlepoint approximation performs particularly badly in comparison to the Hermite approximation. We compare the relative accuracies of the two approximations using the following statistic:
\begin{equation}
B=\frac{\int_0^{\infty}\left|f(x)-\mbox{Saddle}(x)\right|dx}{\int_0^{\infty}\left|f(x)-\mbox{Hermite}(x)\right|dx}.
\label{B}
\end{equation}
where $f(x)$ represents the true distribution and $\mbox{Hermite}(x)$ and $\mbox{Saddle}(x)$ represent the Hermite approximation and saddlepoint approximation respectively. If the saddlepoint approximation is more accurate than the Hermite approximation, one would expect $B$ to be less than 1.

First, for the CIR process, we choose to simulate 100 time series with the parameter values:
\[
	b=0.12,\ \mu=0.05,\ \sigma=0.02,\ x_0=0.049,\ \Delta t=1/52.
\]
For the above parameter values, the relative efficiency statistic, $B=27.96$. Second, for Geometric Brownian motion, we choose to simulate 100 time series with the parameter values:
\[
	\mu=0.12,\ \sigma=0.2,\ x_0=0.049,\ \Delta t=1/12.
\]
For the above parameter values, $B=2.99$. For both diffusion processes, the saddlepoint approximation to the transitional distribution is considerably less accurate than the Hermite approximation. 

Table \ref{coverage} shows the respective coverages of the saddlepoint MCMC, the two Hermite MCMCs (in the table, the traditional algorithm is referred to as {\em Hermite} whilst the modified algorithm developed by \citet{Stramer10} is referred to as {\em Hermite*}) and the true MCMC for both the CIR process as well as Geometric Brownian Motion. A Euler-Maruyama scheme was used to determine the proposal density for the external variates of the {\em Hermite*} procedure. Despite running the comparison for both models at points where the saddlepoint is less accurate than the Hermite approximation, the credibility intervals of the saddlepoint MCMC is closer to the true credibility intervals. The results indicate that the saddlepoint MCMC may be reliably used both to estimate the parameters of a diffusion model as well as to derive their corresponding credibility intervals. This suggests that the stability of the approximations {\em{throughout}} the parameter space is more important than their accuracy at the true parameter values. The credibility intervals produced by the traditional Hermite MCMC have severe undercoverage. This is particularly true for the Geometric Brownian motion since the MCMC chains were started far from the true parameter values ($\mu_0=0.3$, $\sigma_0=0.1$) and the chains often became stuck. For the CIR process, all the chains were started at the true parameter values ($b_0=0.12$, $\mu_0=0.05$, $\sigma_0=0.02$). This serves to highlight the convergence problems that the Hermite MCMC chains suffer from. Though, the modified Hermite* MCMC improves the coverage of the credibility intervals, there is still evidence of undercoverage for the Geometric Brownian motion. This seems to be due to the modified MCMC sometimes taking longer than the burn-in period to converge rather than the chains getting stuck. This is a topic for future research.
\begin{table}
\centering
\linespread{1.5}
\caption{Observed coverage for the four methods for constructing 90\% credibility intervals. {\em Hermite} refers to a traditional MCMC implementation of the Hermite approximation and {\em Hermite*} refers to the modified algorithm developed by \citet{Stramer10}.} 
\begin{tabular}{p{0.3cm} p{0.5cm} p{0.7cm} p{0.95cm} p{1cm} p{0.01cm} p{0.3cm} p{0.5cm} p{0.7cm} p{0.95cm} p{1.2cm}}
\hline
\multicolumn{5}{c}{CIR process} && \multicolumn{5}{c}{Geometric Brownian} \\
\hline
Par. & True & Saddle & Hermite & Hermite* && Par. & True & Saddle & Hermite & Hermite* \\
\hline
$b$ & 0.81 & 0.79 & 0.61 & 0.70 && $\mu$ & 0.90 & 0.90 & 0.14 & 0.44\\
$\mu$ & 0.82 & 0.79 & 0.60 & 0.81 && $\sigma$ & 0.90 & 0.93 & 0.14 & 0.48\\
$\sigma$ & 0.88 & 0.86 & 0.66 & 0.84 && & & & & \\
\hline
\end{tabular}
\label{coverage}
\end{table}

Table \ref{statistics} shows the time taken to study the coverage properties of the credibility intervals as well as the average acceptance ratios for the proposed MCMC steps. There are indications that the acceptance ratios for the Hermite procedures are lower. Both the saddlepoint and the Hermite approximations schemes take considerably longer to run than the case where the true distribution is known. The Saddle MCMC takes 70\% longer to run than the traditional Hermite for the CIR process and 217\% longer for the Geometric Brownian motion. However, the modifications to the traditional algorithm, Hermite* MCMC take considerably longer to run. Hermite* MCMC takes 81\% longer than the Saddle MCMC for the CIR process and 6\% longer for the Geometric Brownian motion.
\begin{table}
\centering
\linespread{1.5}
\caption{Time (in hours) required to run 100 MCMC chains of length 20,000 for the four methods as well as the corresponding average acceptance ratios. The chains were run in MATHEMATICA v6.0 on a Duo Core 2.00 GHz CPU with 4GB RAM. } 
\begin{tabular}{p{1cm} p{0.5cm} p{0.7cm} p{0.95cm} p{1cm} p{0.01cm} p{0.5cm} p{0.7cm} p{0.95cm} p{1.2cm}}
\hline
&\multicolumn{4}{c}{CIR process} && \multicolumn{4}{c}{Geometric Brownian} \\
\hline
 & True & Saddle & Hermite & Hermite* &&  True & Saddle & Hermite & Hermite* \\
\hline
time & 9.80 & 102.30 & 60.35 & 184.76 && 1.60 & 98.29 & 31.04 & 104.22\\
ratio & 0.63 & 0.69 & 0.49 & 0.54 && 0.73 & 0.55 & 0.23 & 0.37\\
\hline
\end{tabular}
\label{statistics}
\end{table}

\subsection{Coverage for non-standard diffusion processes}
We extend our study of the coverage of the saddlepoint procedure to a nonlinear, multivariate diffusion process with unknown transitional distribution. Consider a bivariate-system that behaves as follows:
\begin{eqnarray}
	d\phi_1&=&\left(a\phi_1\phi_2-b\phi_1^2\right)dt+c\phi_2dB^{(1)}_t\nonumber\\
	d\phi_2&=&g\left(\phi^*-\phi_2\right)+\sigma dB^{(2)}_t,
	\label{nl}
\end{eqnarray}
where the Brownian motions are independent. Both processes are mean-reverting: $\phi_2$ is an Ornstein-Uhlenbeck process and for a fixed value of $\phi_2$, the process $\phi_1$ will have long-term mean $a\phi_2/b$. As $\phi_2$ increases, the instantaneous mean and variance of $\phi_1$ also increases. From equations (\ref{MGF}) and (\ref{rel}), the corresponding partial differential for $K(\nu_1,\nu_2,t)=\ln \mbox{E}[\exp\{\nu_1\phi_1+\nu_2\phi_2\}]$ is given by:
\begin{small}
\begin{eqnarray}
	\frac{\partial K}{\partial t}&=&a\nu_1\left[\frac{\partial K}{\partial\nu_1}\frac{\partial K}{\partial\nu_2}+\frac{\partial^2K}{\partial\nu_1\partial\nu_2}\right]-b\nu_1\left[\left(\frac{\partial K}{\partial\nu_1}\right)^2+\frac{\partial^2K}{\partial\nu_1^2}\right]\nonumber\\
	&+&\left[\frac{c\nu_1^2}{2}-g\nu_2\right]\frac{\partial K}{\partial\nu_2}+g\phi^*\nu_2+\frac{\sigma^2}{2}\nu_2^2.
	\label{nlCGF}
\end{eqnarray}
\end{small}
A third order cumulant truncation of $K(\nu_1,\nu_2,t)$ yields:
\begin{eqnarray}
	K(\nu_1,\nu_2,t)&\approx&1+\nu_1\kappa_{10}(t)+\nu_2\kappa_{01}(t)+\nu_1\nu_2\kappa_{11}(t)\nonumber\\
	&+&\frac{\nu_1^2}{2!}\kappa_{20}(t)+\frac{\nu_2^2}{2!}\kappa_{02}(t).
	\label{biExp}
\end{eqnarray}
(\ref{biExp}) may be substituted within (\ref{nlCGF}) to give a system of ordinary differential equations for the cumlulants that can subsequently be solved and used within the saddlepoint approximation.

The diffusion system has six parameters: $a$, $b$, $c$, $g$, $\phi^*$ and $\sigma$. Suppose from (\ref{nl}) we simulate time series, each of length 100, for $\phi_1$ and $\phi_2$. As in the previous subsection, an investigation into the coverage of the credibility intervals is computationally expensive and hence is only feasible for a single set of parameter values. We simulate 100 time series using the parameter values:
\[
	a=0.1, b=0.02, c=1.8, g=0.5, \phi^*=5, \sigma=1.
\]
For each time series, the saddlepoint procedure is used to construct 90\% credibility intervals for the six parameters. The observed coverages for the parameters are:
\[
	a: \frac{80}{100};\mbox{\ \ } b: \frac{82}{100};\mbox{\ \ } c: \frac{83}{100};\mbox{\ \ } g: \frac{87}{100};\mbox{\ \ } \phi^*: \frac{82}{100};\mbox{\ \ } \sigma: \frac{81}{100}.
\]
These coverages are close to the advertised values, suggesting that the procedure works well in this nonlinear, multivariate example.
\section{Application to Financial Data}
\label{sec:Ex}
\begin{figure}[ht]
	\caption{S\&P 500 and VIX indices from December 2009 to November 2010. The left y-axis is for the S\&P and the right y-axis is for the VIX index. The VIX index has been scaled to ensure that the Heston model may mimic the variance of the S\&P}
	\begin{center}
		\includegraphics[width=4.3in]{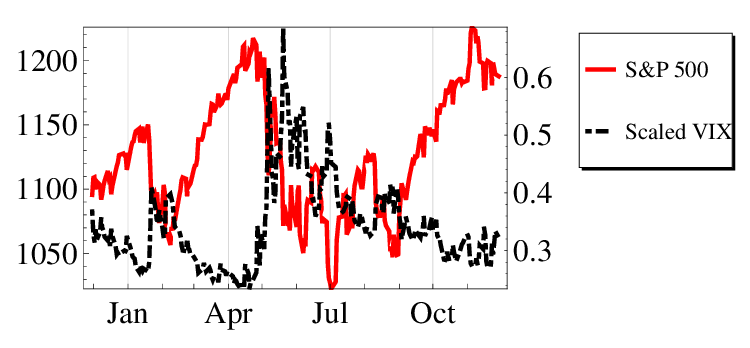}
	\end{center}
	\label{data}
\end{figure}

We study the S\&P 500 and it's relation to the VIX index (a popular measure of implied volatility) over the period December 2009 to November 2010. Figure \ref{data} shows the two indices over this period. The Heston model is fitted to the dataset. It uses proxies for market volatility - such as the VIX index - to account for market movements. The key feature of this model is that the volatility of the asset price movements are assumed to be stochastic.
\begin{equation}
	d\left[
\begin{array}{c}
	S_t\\
	V_t
\end{array}\right]
=\left[\begin{array}{c}
	rS_t\\
	\delta(\theta-V_t)
\end{array}\right] dt
+\left[
\begin{array}{cc}
	S_t\sqrt{(1-\rho^2)V_t} & \rho S_t\sqrt{V_t}\\
	0 & \sigma \sqrt{V_t}
\end{array}\right]
d\left[
\begin{array}{c}
	B_t^{(1)}\\
	B_t^{(2)}
\end{array}
\right],
\label{Heston}
\end{equation}
where $S_t$ represents the asset price, $V_t$ is the underlying volatility that drives the asset movements and $r$, $\delta$, $\theta$, $\rho$ and $\sigma$ are parameters. The Heston model is a multivariate, irreducible diffusion processes for which many of the competing methods would not be applicable.

We wish to run the saddle MCMC in order to obtain both estimates and credibility intervals for the parameters for the Heston model. To run the saddle MCMC, the diffusion tensor $\sigma_{ij}(S_t,V_t;\theta)$ is needed. This is given by:
\begin{footnotesize}
\begin{equation}
\left[
\begin{array}[pos]{cc}
	S_t\sqrt{(1-\rho^2)V_t} & 0\\
	\rho S_t\sqrt{V_t} & \sigma \sqrt{V_t}
\end{array}\right]^T
\left[
\begin{array}[pos]{cc}
	S_t\sqrt{(1-\rho^2)V_t} & 0\\
	\rho S_t\sqrt{V_t} & \sigma \sqrt{V_t}
\end{array}\right]
=\left[
\begin{array}[pos]{cc}
	S_t^2 V_t & \rho\sigma V_tS_t\\
	\rho\sigma V_tS_t & \sigma^2 V_t
\end{array}\right].\nonumber
\end{equation}
\end{footnotesize}
By applying equations (\ref{MGF}) and (\ref{rel}) to the above diffusion process, we obtain the following partial differential equation for $K(\nu_1,\nu_2,t)=\ln \mbox{E}[\exp\{\nu_1S_t+\nu_2V_t\}]$:

\begin{small}
\begin{align}
\frac{\partial K(\nu_1,\nu_2,t)}{\partial t}&=\nu_1 r \frac{\partial K}{\partial \nu_1}+\nu_2\delta\theta-\nu_2\delta\frac{\partial K}{\partial \nu_2}\nonumber\\
&+\frac{1}{2}\nu_1^2\left(\frac{\partial K}{\partial \nu_1}\right)^2\left(\frac{\partial K}{\partial \nu_2}\right)+\nu_1^2\left(\frac{\partial K}{\partial \nu_1}\right)\left(\frac{\partial^2 K}{\partial\nu_1\partial\nu_2}\right)\nonumber\\
&+\frac{1}{2}\nu_1^2\left(\frac{\partial^2K}{\partial\nu_1^2}\right)\left(\frac{\partial K}{\partial\nu_2}\right)+\frac{1}{2}\nu_1^2\frac{\partial^3K}{\partial \nu_1^2\partial\nu_2}\nonumber\\
&+\frac{1}{2}\nu_2^2\sigma^2\frac{\partial K}{\partial\nu_2}+\nu_1\nu_2\rho\sigma\left(\frac{\partial K}{\partial\nu_1}\right)\left(\frac{\partial K}{\partial\nu_2}\right)\nonumber\\
&+\nu_1\nu_2\rho\sigma\frac{\partial^2K}{\partial\nu_1\partial\nu_2}.
\end{align}
\end{small}
A third order cumulant truncation is performed to yield a system of ordinary differential equations for the cumulants. These may be subsequently used to approximate the likelihood of the Heston model. The modified MCMC (described in section \ref{sec:MCMC}) may then be run on the S\&P and scaled VIX dataset.

The resulting parameter estimates and 95\% credibility intervals are shown in Table \ref{estima}. One may see that there is strong negative correlation between the two Brownian motions driving $S_t$ and $V_t$. The parameters $r$ and $\delta$ have wide credibility intervals, suggesting that they are not strongly constrained by the data.
\begin{table}
\centering
\linespread{1.5}
\caption{Parameter estimates for the S\&P and scaled VIX indices over the period December 2009 to November 2010.}
\begin{tabular}{c c c}
\hline
Par. & Estimate & 95\% C.I. \\
\hline
$r$ & 0.118 & (0.008; 0.430) \\
$\delta$ & 9.863 & (1.410; 13.464) \\
$\theta$ & 0.034 & (0.020; 0.058) \\
$\rho$ & -0.855 & (-0.880; -0.824) \\
$\sigma$ & 0.250 & (0.234; 0.266)\\
\hline
\end{tabular}
\label{estima}
\end{table}

\section{Conclusion}
A general statistical framework for estimating the parameters of a diffusion system is proposed. It is assumed that the elements of both the mean vector and the diffusion matrix may be represented by finite order polynomials. Apart from this, the method may be readily applied to multivariate diffusion systems. Many existing parameter estimation methods are only applicable for univariate diffusion processes or, if applicable to multivariate systems, require the diffusion system to be reducible \citep{Ait2008,Beskos06}.

The transitional distribution of a diffusion system is approximated with a saddlepoint. We show how the cumulants of the system may be predicted and subsequently used to approximate the true transitional distribution of the system. The proposed method is compared with the Hermite approximation proposed by \citet{Ait2002}. For both the Cox-Ingersoll Ross model and Geometric Brownian motion, the accuracy of the saddlepoint is comparable to the Hermite approximation at the true parameter values. Furthermore, since the Hermite approximation is a small-time expansion, the Hermite approximation rapidly becomes inaccurate as the time interval between successive observations increases.

The saddlepoint approximations may be easily incorporated within an MCMC algorithm. Since, in regions far from the maximum likelihood estimates, the saddlepoint approximation to the transitional density is better behaved than the Hermite approximation, the saddlepoint MCMC chains have better convergence properties. Consequently, the coverage of the credibility intervals for the saddlepoint MCMC is close to those obtained from the true MCMC. In contrast, the credibility intervals for the unmodified Hermite MCMC suffer from undercoverage. This undercoverage is especially severe when the chains are not started at the true parameter values since the chains often become stuck. This is mitigated by the modified algorithm developed by \citet{Stramer10} which does not require the computation of the posterior's normalizing constant. However, the modified chains are not only slower than the saddlepoint MCMC, but they take longer to converge, which caused undercoverage for the Geometric Brownian motion example studied in Section \ref{sec:Comp} since the chains had not converged by the end of the burn-in period. 

The comparison between the saddlepoint and Hermite approximations was only performed for univariate diffusion processes and hence caution must be exercised in making claims for multivariate diffusion processes. However, the good coverage observed for the credibility intervals of the saddlepoint MCMC for the non-linear, multivariate example suggests that the procedure still works well for multivariate diffusion processes. Also, for irreducible multivariate processes, the Hermite procedure requires a two-fold approximation, which leads to a further drop in its accuracy \citep{Ait2008}.


The saddlepoint MCMC is used to fit the Heston model to the S\&P 500 and the VIX indices over the period December 2009 to November 2010. Since the Heston model is a multivariate, irreducible diffusion process, the estimation of the model parameters presents a formidable challenge, but the saddlepoint method provides reliable estimates and credibility intervals for the model parameters. As expected, there was significant negative correlation between the two Brownian motions driving the asset prices and the volatility process.


The proposed estimation algorithm has several virtues: it is fast; applicable for a wide range of diffusion systems and gives parameter estimates close to the true values. Furthermore, the corresponding credibility intervals for these estimates have coverage close to their advertised values. This makes the algorithm suitable for estimating the parameters of many diffusion systems.

\section*{Acknowledgments}
\noindent This material is based upon work supported financially by the National Research Foundation of South Africa. The author is grateful to two referees  for their constructive comments which have greatly improved the manuscript. Thanks are also due to Drs Trevor Hastie, Leonard Stefanski and Eric Renshaw for helpful discussions. Part of this work was performed whilst hosted by the Department of Mathematics and Statistics at the University of Melbourne.

\section*{Appendix: Derivation of equation (\ref{MGF})}
\label{app:scripts}
The MGF of a diffusion system behaves according to the following partial differential equation:
\begin{equation}
\frac{\partial M(\Lambda,t)}{\partial t}=\bigg[\sum_{i=1}^m \upsilon_i\mu_i(\Lambda^*,t)
+\frac{1}{2}\sum_{i=1}^m\sum_{j=1}^m \upsilon_i\upsilon_j\sigma_{ij}(\Lambda^*,t)\bigg]M(\Lambda,t).
\label{appMGF}
\tag{A.1}
\end{equation}
{\bf{Proof:}} Consider the multivariate diffusion system $\phi_t^*$ with instantaneous mean vector $\mu(\phi_t^*,t)$ and diffusion matrix $\sigma(\phi_t^*,t)$. We denote the transition probability by:
\[
Pr\left[\Delta \phi_t^*=\Delta x | \phi_t^*\right]=g\left(\Delta x|\phi_t^*\right).
\]
Let $\Lambda^T=(\upsilon_1,...,\upsilon_m)$ be the parameters of the MGF and let $\Lambda^*=(\partial / \partial\upsilon_1,..., \partial / \partial\upsilon_m)$. The MGF,  $M(\Lambda,t)=E\left[\exp[\upsilon_1\phi_1(t)+...+\upsilon_m\phi_m(t)]\right]$ obeys the following differential equation \citep{Barbour72}:
\begin{equation}
\frac{\partial M(\Lambda,t)}{\partial t}=\sum_{\Delta x}\left(e^{\Lambda^T\Delta x}-1\right)g\left(\Delta x|\Lambda^*\right)M(\Lambda,t).
\label{Barb}
\tag{A.2}
\end{equation}
For small $\Delta x$, we have:
\begin{equation}
e^{\Lambda^T\Delta x}-1=e^{\sum\upsilon_i\Delta x_i}-1
=\sum_{i=1}^m \upsilon_i\Delta x_i +\frac{1}{2} \sum_{i=1}^m\sum_{j=1}^m\upsilon_i\upsilon_j\Delta x_i\Delta x_j. \nonumber
\label{ser}
\tag{A.3}
\end{equation}
For diffusion processes the higher order terms are negligible. We thus only require the joint and marginal transitional rates of the diffusion system to characterize the evolution of the MGF. We denote the joint transition probabilities for $\phi_i$ and $\phi_j$ by:
\[
Pr\left[\Delta\phi_i=x_i, \Delta\phi_j=x_j|\phi_t^*\right]=g_{ij}\left(x_i,x_j|\phi_t^*\right).
\]
The marginal transition rate is denoted by:
\[
Pr\left[\Delta\phi_i=x_i|\phi_t^*\right]=g_i\left(x_i|\phi_t^*\right).
\]
Note that the instantaneous mean $\mu_i(\phi_t^*,t)$ and covariances $\sigma_i(\phi_t^*,t)$ can be represented in terms of the joint and marginal transitional rates:
\begin{equation}
\mu_i(\phi_t^*,t)=\lim_{\Delta t, \Delta x_i\rightarrow0} \frac{\Delta x_i}{\Delta t} \left(g_i\left(\Delta x_i|\phi_t^*\right)-g_i\left(-\Delta x_i|\phi_t^*\right)\right).
\tag{A.4}
\end{equation}
\begin{small}
\begin{multline}
\sigma_{ij}(\phi_t^*,t)=\lim_{\Delta t, \Delta x_i,\Delta x_j\rightarrow0} \frac{(\Delta x_i\Delta x_j)}{\Delta t} \bigg(g_{ij}\left(\Delta x_i,\Delta x_j|\phi_t^*\right)\\
+g_{ij}\left(-\Delta x_i,-\Delta x_j|\phi_t^*\right)-g_{ij}\left(\Delta x_i,-\Delta x_j|\phi_t^*\right)\\
-g_{ij}\left(-\Delta x_i,\Delta x_j|\phi_t^*\right)\bigg).
\tag{A.5}
\end{multline}
\end{small}
By substituting \eqref{ser} within \eqref{Barb}, we obtain:
\begin{small}
\begin{multline}
\frac{\partial M(\Lambda,t)}{\partial t}=\sum_{i=1}^m \Delta x_i \left[\frac{g_i(\Delta x_i|\Lambda^*)}{\Delta t}+\frac{g_i(-\Delta x_i|\Lambda^*)}{\Delta t}\right]\upsilon_i M\\
+\frac{1}{2}\sum_{i=1}^m\sum_{j=1}^m\Delta x_i \Delta x_j \bigg[\frac{g_{ij}(\Delta x_i,\Delta x_j|\Lambda^*)+g_{ij}(-\Delta x_i,-\Delta x_j|\Lambda^*)}{\Delta t} \\
-\frac{g_{ij}(\Delta x_i,-\Delta x_j|\Lambda^*)+g_{ij}(-\Delta x_i,\Delta x_j|\Lambda^*)}{\Delta t}\bigg].
\label{close}
\tag{A.6}
\end{multline}
\end{small}
The expressions for the instantaneous mean and variance may be substituted within (\ref{close}). This gives us:
\begin{equation}
\frac{\partial M(\Lambda,t)}{\partial t}=\bigg[\sum_{i=1}^m \upsilon_i\mu_i(\Lambda^*,t)
+\frac{1}{2}\sum_{i=1}^m\sum_{j=1}^m \upsilon_i\upsilon_j\sigma_{ij}(\Lambda^*,t)\bigg]M(\Lambda,t).
\tag{A.7}
\end{equation}

\bibliographystyle{elsart-harv}
\bibliography{references}





\end{document}